\documentclass{article}
\usepackage[utf8]{inputenc}
\usepackage[T1]{fontenc}

\usepackage{amssymb} 
\usepackage{amsmath}

\usepackage{amsthm}

\newtheorem{theorem}{Theorem}
\newtheorem{definition}{Definition}
\usepackage[affil-it]{authblk}

\usepackage{hyperref}

\usepackage{listings}
\usepackage{proof}

\usepackage[inline]{enumitem} 

\usepackage{makecell}

\newcommand{\comment}[1]{}

\newcommand{\state}{\mathbf{State}}
\newcommand{\cartprodstate}[0]{\state \times \state}
\newcommand{\powerstate}[0]{2^{\cartprodstate}}
\newcommand{\trace}[0]{\state^{+}}
\newcommand{\defeq}{\stackrel{\mathsf{def}}{=}}

\newcommand{\set}[1]{\left\{ #1 \right\}}
\newcommand{\setdef}[2]{\left\{ #1 \mid #2 \right\}}
\newcommand{\denot}[1]{\left[\!\left[#1\right]\!\right]}

\newcommand{\env}{\mathbf{Env}}
\newcommand{\fenv}[1]{\env_{#1}} 
\newcommand{\fenvtype}[1]{\fenv{#1} = #1 \rightarrow \powerstate}
\newcommand{\fprov}[1]{P^+_{#1}}
\newcommand{\freq}[1]{P^-_{#1}}
\newcommand{\anyprov}[2]{{#1}^+_{#2}}
\newcommand{\anyreq}[2]{{#1}^-_{#2}}
\newcommand{\funtype}[1]{\fenv{\freq{#1}} \rightarrow \fenv{\fprov{#1}}}
\newcommand{\comptype}[1]{#1 : \funtype{#1}}

\newcommand{\intfpair}[1]{(\freq{#1}, \fprov{#1})}

\newcommand{\intfsym}[1]{I_{#1}}
\newcommand{\intfeq}[1]{\intfsym{#1} = \intfpair{#1}}

\newcommand{\mcont}[1]{\mathcal{C}_{#1}}
\newcommand{\menv}[1]{E_{#1}}
\newcommand{\mimp}[1]{M_{#1}}
\newcommand{\mpair}[1]{(\menv{#1}, \mimp{#1})}

\newcommand{\hoare}[3]{\ensuremath{\{#1\}~#2~\{#3\}}}
\newcommand{\ruleName}[1]{\textsc{#1}}

\newcommand{\todo}[1]{\textbf{[#1]}}

\newcommand{\excomp}{{m_\mathit{even} \times m_\mathit{odd}}}

\title{An Abstract Contract Theory for \\ Programs with Procedures\\
\sc{Full Version}\thanks{This is the full version of the paper
\emph{An Abstract Contract Theory for Programs with
Procedures}~\cite{lid-gur-21},
published in \emph{Proceedings of the 24th International Conference 
on Fundamental Approaches to Software Engineering} (FASE 2021),
which includes the proofs of all theorems and additional examples.
The conference version should always be cited.}
}
\author{Christian Lidström}
\author{Dilian Gurov}
\affil{KTH Royal Institute of Technology, Stockholm, Sweden}

\begin{document}

\maketitle

\begin{abstract}
When developing complex software and systems,
\emph{contracts} provide a means for controlling the
complexity by dividing the responsibilities among the
components of the system in a hierarchical fashion.
In specific application areas, dedicated \emph{contract theories}
formalise the notion of contract and the operations
on contracts in a manner that supports best the development
of systems in that area.
At the other end, \emph{contract meta-theories} attempt to
provide a systematic view on the various contract theories
by axiomatising their desired properties.
However, there exists a noticeable gap between the most
well-known contract meta-theory of Benveniste
et al.~\cite{ben-et-al-18}, which focuses on the
design of embedded and cyber-physical systems,
and the established way of using contracts when developing
general software, following Meyer's design-by-contract
methodology~\cite{meyer-92}.
At the core of this gap appears to be the notion of
\emph{procedure}: while it is a central unit of composition in
software development, the meta-theory does not suggest
an obvious way of treating procedures as components.

~~~In this paper, we provide a first step towards a contract
theory that takes procedures as the basic building block, and is
at the same time an instantiation of the meta-theory.
To this end, we propose an abstract contract theory for
sequential programming languages with procedures,
based on \emph{denotational semantics}.
We show that, on the one hand, the specification of contracts
of procedures in \emph{Hoare logic}, and their procedure-modular
verification, can be cast naturally in the framework of our abstract
contract theory.
On the other hand, we also show our contract theory to fulfil
the axioms of the meta-theory.
In this way, we give further evidence for the utility of the
meta-theory, and prepare the ground for combining our instantiation
with other, already existing instantiations.
\end{abstract}

\pagestyle{plain}


\section{Introduction}

\paragraph{Contracts.}

Loosely speaking, a \emph{contract} for a software or system
component is a means of specifying that the component obliges 
itself to guarantee a certain behaviour or result, provided that 
the user (or client) of the component obliges itself to fulfil 
certain constraints on how it interacts with the component. 

One of the earliest inspirations for the notion of 
software contracts came 
from the works of Floyd~\cite{floyd-67} and
Hoare~\cite{hoare-69}. One outcome of this was
\emph{Hoare logic}, which is a way of assigning meaning
to sequential programs \emph{axiomatically}, through 
so-called Hoare triples. 
A Hoare triple $\{P\} S \{Q\}$ consists of two
assertions $P$ and $Q$ over the program variables,
called the pre-condition and post-condition,
respectively, and a program $S$.
The triple states that if the pre-condition~$P$
holds prior to executing~$S$, then, if execution of~$S$ 
terminates, the post-condition~$Q$ will hold upon termination.
With the help of additional, so-called \emph{logical variables},
one can specify, with a Hoare triple, the desired
relationship between the final values of certain
variables (such as the return value of a procedure) 
and the initial values of certain other variables 
(such as the formal parameters of the procedure).

This style of specifying contracts has been advocated
by Meyer~\cite{meyer-92}, together with the
design methodology Design-by-Contract.
A central characteristic of this methodology is that it
is well-suited for \emph{independent implementation and 
verification}, where software components are developed
independently from each other, based solely on the
contracts, and without any knowledge of the
implementation details of the other components.

\paragraph{Contract Theories.}

Since then, many other contract theories have emerged, 
such as Rely/Guarantee reasoning~\cite{jones-83,van-staden-15}
and a  number of Assume/Guarantee contract
theories~\cite{ben-et-al-07,benvenuti-et-al-08}.
A contract theory typically formalises the notion of contract,
and develops a number of operations on contracts that support 
typical design steps.
This in turn has lead to a few developments of contract
\emph{meta-theories}
(e.g. \cite{ben-et-al-18,bau-et-al-12,chen-et-al-12}),
which aim at unifying these,
in many cases incompatible, contract theories.
The most comprehensive, and well-known, of these,
is presented in Benveniste et al.~\cite{ben-et-al-18},
and is concerned specifically with the design of
cyber-physical systems.
Here, all properties
are derived from a most abstract notion of a contract. 
The meta-theory focuses on the notion of contract 
\emph{refinement}, and the operations of contract 
\emph{conjunction} and \emph{composition}. 
The intention behind refinement and composition is to 
support a top-down design flow, where contracts are 
decomposed iteratively
into sub-contracts; the task is then to show that
the composition of the sub-contracts refines the
original contract. 
These operations are meant to enable
\emph{independent development}
and \emph{reuse} of components.
In addition, the operation of conjunction is
intended to allow the superimposition of contracts
over \emph{the same} component, when they concern 
different aspects of its behaviour.
This also enables \emph{component reuse}, by
allowing contracts to reveal only the behaviour
relevant to the different use cases.

\paragraph{Motivation and Contribution.}

The meta-theory of Benveniste et al.\ focuses on the
design of embedded and cyber-physical systems.
However, there exists a noticeable gap between this
meta-theory 
and the way contracts are used when developing 
general software following Meyer's design-by-contract 
methodology. 
At the core of this gap appears to be the notion of 
\emph{procedure}\footnote{We use the term ``procedure'', 
rather than ``function'' or ``method'',
to refer to the well-known control abstraction mechanism 
of imperative programming languages.
}.
While the procedure is a central unit of composition 
in software development, the meta-theory does not suggest
an obvious way of treating procedures as components. 
This situation is not fully satisfactory, since the software
components of most embedded systems are implemented with
the help of procedures (a typical C-module, for instance,
would consist of a main function and a number of \emph{helper}
functions), and their development should ideally follow 
the same design flow as that of the embedded system as a whole. 

In this paper we provide a first step towards a contract
theory that takes procedures as the basic building block,
and at the same time respects the axioms of the meta-theory.
Our contract theory is abstract,
so that it can be instantiated to any procedural language,
and similarly to the meta-theory, 
is presented at the semantics level only. Then,
in the context of a simplistic imperative programming language with 
procedures and its denotational semantics, 
we show that the specification of contracts of procedures 
in \emph{Hoare logic}, and their procedure-modular
verification, can be cast in the framework of our abstract 
contract theory. We also show that our contract theory 
is an instance of the meta-theory of Benveniste et al.
With this we expect to contribute to the bridging of the
gap mentioned above, and to give a formal
justification of the design methodology supported by the
meta-theory, when applied to the software components of
embedded systems.
Several existing contract theories have already
been shown to instantiate the meta-theory.
In providing a contract theory for procedural programs that
also instantiates it, we increase the 
value of the meta-theory by providing further evidence for its 
universality. 
In addition, we prepare the theoretical ground for combining 
our instantiation with other instantiations, which may target 
components not to be implemented in software. 

Our theoretical development should be seen as a proof-of-concept. 
In future work it will need to be extended to cover more 
programming language features, such as object orientation,
multi-threading, and exceptions. 

\paragraph{Related Work.}

Software contracts and operations on contracts have long been
an area of intensive research, as evidenced, e.g., 
by~\cite{aba-lam-93}.
We briefly mention some works related to our theory, in 
addition to the already mentioned ones. 

Reasoning from multiple Hoare triples is studied
in~\cite{owe-et-al-17},
in the context of unavailable source code, where
new properties cannot be derived by re-verification.
In particular, it is found that two Hoare-style
rules, the standard rule of consequence and a generalised
normalisation rule, are sufficient to infer,
from a set of existing contracts for a procedure,
any contract that is semantically entailed. 

Often-changing source code is a problem for
contract-based reasoning and contract reuse.
In~\cite{hahnle-et-al-13}, abstract method calls
are introduced to alleviate this problem.
Fully abstract contracts are then introduced
in~\cite{bub-et-al-14}, allowing reasoning about software
to be decoupled from contract applicability checks, in a
way that not all verification effort is invalidated by
changes in a specification.

The relation between behavioural specifications and
assume/guarantee-style contracts for modal transition
systems is studied
in~\cite{bau-et-al-12}, which shows how to build a
contract framework from any specification theory supporting
composition and refinement.
This work is built on in~\cite{cim-ton-15}, where a formal
contract framework based on temporal logic is presented,
allowing verification of correctness of contract refinement
relative to a specific decomposition.

A survey of behavioural specification 
languages~\cite{hat-et-al-12}
found that existing languages are well-suited
for expressing properties of software components, but it
is a challenge to express how components interact,
making it difficult to reason about system and architectural
level properties from detailed design specifications.
This provides additional evidence for the gap between
contracts used in software verification and contracts
as used in system design.

\paragraph{Structure.}

The paper is organised as follows. 
Section~\ref{sec:contract-based-design} recalls the
concept of contract based design and the
contract meta-theory considered in the present paper.
In Section~\ref{sec:preliminaries} we present a denotational
semantics for programs with procedures, including a semantics
for contracts for use in procedure-modular verification.
%
Next, Section~\ref{sec:proc-lang-contract-theory} presents our
abstract contract theory for sequential programs with procedures.
Then, we show in
Section~\ref{sec:meta-connection} that our contract theory 
fulfils the axioms of the meta-theory, while in
Section~\ref{sec:hoare-connection} we show
how the specification of contracts of procedures 
in Hoare logic and their procedure-modular verification can be 
cast in the framework of our abstract contract theory. 
We conclude with Section~\ref{sec:conclusion}.


\section{Contract Based Design}
\label{sec:contract-based-design}

This section describes the concept of
\emph{contract based design}, and motivates its
use in cyber-physical systems development.
We then recall the contract meta-theory by
Benveniste et al.~\cite{ben-et-al-18}.


\subsection{Contract Based Design of Cyber-Physical Systems}
\label{subsec:cbd-of-cps}

\emph{Contract based design} is an approach to systems design,
where the system is developed in a top-down manner through the
use of contracts for components, which are incrementally assembled
so that they preserve the desired system-wide properties.
Contracts are typically described
by a set of \emph{assumptions} the component makes on its environment,
and a set of \emph{guarantees} on the component's behaviour, given
that it operates in an environment adhering to the
assumptions~\cite{ben-et-al-18}.

Present-day cyber-physical systems, such as those found in
the automotive, avionics and other industries, are extremely
complex. Products assembled by Original Equipment Manufacturers
(OEMs) often consist of components from a number of different
suppliers, all using their own specialised design processes,
system architectures, development platforms, and tools.
This is also true inside the OEMs, where there are different
teams with different viewpoints of the system, and their own
design processes and tools. In addition, the system itself has
several different aspects that need to be managed, such as
the architecture, safety and security requirements, functional
behaviour, and so on.
Thus, a rigorous design framework is called for that can solve
these design-chain management issues.

Contract based design addresses these challenges through the
principles, at the specification level,
of \emph{refinement} and \emph{abstraction},
which are processes for managing the design flow between
different layers of abstraction, and \emph{composition}
and \emph{decomposition}, which manage the flow at
the same level of abstraction.
Generally, when designing a system, at the top level
of abstraction there will be an overall system specification
(or contract).
This \emph{top-level contract} is then
refined, to provide a more concrete contract for the
system, and decomposed, in order to obtain contracts
for the sub-systems, and to separate the
different viewpoints of the system.
A system design typically iterates the
decomposition-and-refinement process, resulting in several
layers of abstraction, until contracts are obtained that
can be directly implemented, or for which implementations
already exist. 
An important requirement on this methodology of
hierarchical decomposition and refinement of
contracts is that it must guarantee
that when the low-level components
implement their concrete contracts, and are combined to
form the overall system, then the top-level, abstract,
contract shall hold.

Furthermore, a contract framework in particular needs
to support \emph{independent development} and
\emph{component reuse}. That is, specifications
for components, and their operations, must allow for
components and specifications to be independently designed
and implemented, and to be used in different parts of the
system, each with their own assumptions on how the other
components, the environment, behave.
This is achieved through the principle operations on
contracts: \emph{refinement}, \emph{composition},
and \emph{conjunction}.

Refinement allows one to extract a contract at the
appropriate level of abstraction.
A desired property of refinement is that components
which have been designed with reference to the more
abstract (i.e., weaker) contract do not need to be
re-designed after the refinement step.
That is, in the early stages of development
an OEM may have provided a weak contract
for some subsystem to an external supplier, which
implemented a component relying on this contract.
As development of the system progresses, and the contract
is refined, the component supplied externally should
still operate according to its guarantees without
needing to be changed, when instead assuming
the new, refined, contract.


Composition enables one to combine contracts of
different components into a contract for the larger
subsystem obtained when combining the components.
Again, a desirable property is that other components
relying on one or more of the individual contracts,
can, after composition of the contracts,
assume the new contract and still perform its
guarantees, without being re-designed, thus
ensuring that subsystems can be independently
implemented.

Finally, contract conjunction is another way of
combining contracts, but now for the
different viewpoints of a single component.
This allows one to separate a contract into
several different, finer
contracts for the same component, revealing just
enough information for each particular system that
depends on it, so that it can be reused
in different parts of the system, or in entirely
different systems.


\subsection{A Contract Meta-Theory}
\label{sec:meta-theory}

We consider the meta-theory described in~\cite{ben-et-al-18}.
The stated purpose of the meta-theory has been to distil 
the notion of a contract to its essence, 
so that it can be used in system design methodologies
without ambiguities.
In particular, the meta-theory has been developed to
give support for design-chain management, and to allow
\emph{component reuse} and \emph{independent development}.
It has been shown that a number of concrete contract
theories instantiate it, including assume/guarantee-contracts,
synchronous Moore interfaces, and interface theories.
To our knowledge, this is the only meta-theory of
its purpose and scope.


We now present the formal definitions of the concepts 
defined in the meta-theory, and the properties that
they entail.
The meta-theory is defined only in terms of semantics,
and it is up to particular concrete instantiations
to provide a syntax.
%



\paragraph{Components. }

The most basic concept in the meta-theory is that
of a \emph{component}, which represents any concrete
part of the system.
Thus, we have an abstract component universe~$\mathbb{M}$
with components $m \in \mathbb{M}$.
Over pairs of components, we have a \emph{composition} operation~$\times$. 
This operation is partially defined,
and two components~$m_1$ and $m_2$ are called
\emph{composable} when $m_1 \times m_2$ is defined.
In such cases, we call~$m_1$ an \emph{environment}
for~$m_2$, and vice versa. 
In addition, component composition must be both
commutative and associative, in order to ensure
that different components can be combined
in any order.

%
Typically, components are \emph{open}, in the sense
that they contain functionality provided by other 
components, i.e., their environment.
The environment in which a component is to be placed is
often unknown at development time, and although a component
cannot restrict it, it is designed for a certain context.


\paragraph{Contracts.}

In the meta-theory, the notion of \emph{contract}
is defined in terms of sets of components.
The contract universe
$\mathbb{C} \defeq 2^\mathbb{M} \times 2^\mathbb{M}$ 
consists of contracts $\mathcal{C} = (E, M)$,
where $E$ and~$M$
are the sets of \emph{environments} 
and \emph{implementations} of~$\mathcal{C}$, respectively.
Importantly, each pair $(m_1, m_2) \in E \times M$
must be composable. 
This definition is intentionally
abstract. The intuition is that contracts separate the
responsibilities of a component from the expectations
on its environment.
Moreover, contracts are best seen as \emph{weak
specifications} of components: they should expose just
enough information to be adequate for their purpose.

For a component~$m$ and a contract $\mathcal{C} = (E, M)$,
we shall sometimes write  $m \models^E \mathcal{C}$ for
$m \in E$, and $m \models^M \mathcal{C}$ for $m \in M$.
A contract $\mathcal{C}$ is said to be \emph{consistent}
if it has at least one implementation, and
\emph{compatible} if it has at least one environment.
%

\begin{table}[t]
\caption{Properties that hold in theories
that adhere to the meta-theory.}
\centering
\begin{tabular}{ | c | l | }
\hline
\textbf{\#} & \textbf{Property} \\
\hline
1 & \makecell[l]{\emph{Refinement.}
    When $\mathcal{C}_1 \preceq \mathcal{C}_2$,
    every implementation of~$\mathcal{C}_1$ 
    is also an \\ implementation of $\mathcal{C}_2$.} \\
\hline
2 & \makecell[l]{\emph{Shared refinement.}
    Any contract refining $\mathcal{C}_1 \wedge \mathcal{C}_2$
    also refines $\mathcal{C}_1$ and $\mathcal{C}_2$.\\
    Any implementation of $\mathcal{C}_1 \wedge \mathcal{C}_2$
    is a shared implementation of $\mathcal{C}_1$ and $\mathcal{C}_2$.\\
    Any environment for $\mathcal{C}_1$ and $\mathcal{C}_2$
    is an environment for $\mathcal{C}_1 \wedge \mathcal{C}_2$.} \\
\hline
3 & \makecell[l]{\emph{Independent implementability.}
    Compatible contracts can be \\ independentlyimplemented.} \\
\hline
4 & \makecell[l]{\emph{Independent refinement.}
    For all contracts $\mathcal{C}_i$ and $\mathcal{C}_i', i \in I$,
    if $\mathcal{C}_i, i \in I$ \\are compatible
    and $\mathcal{C}_i' \preceq \mathcal{C}_i, i \in I$ hold,
    then $\mathcal{C}_i', i \in I$ are compatible \\ and
    $\bigotimes_{i \in I} \mathcal{C}_i' \preceq
    \bigotimes_{i \in I} \mathcal{C}_i$}\\
\hline
5 & \makecell[l]{\emph{Commutativity, sub-associativity.}
    For any finite sets of contracts $\mathcal{C}_i,$\\$ i = 1, \dots, n$,
    $\mathcal{C}_1 \otimes \mathcal{C}_2 =
    \mathcal{C}_2 \otimes \mathcal{C}_1$ and 
    $\bigotimes_{1 \leq i \leq n} \mathcal{C}_i \preceq
    (\bigotimes_{1 \leq i < n} \mathcal{C}_i)
    \otimes \mathcal{C}_n$ \\ holds.} \\
\hline
6 & \makecell[l]{\emph{Sub-distributivity.}
    The following holds, if all contract compositions\\ in the
    formula are well defined:\\
    $\left( (\mathcal{C}_{11} \wedge \mathcal{C}_{21}) \otimes
        (\mathcal{C}_{12} \wedge \mathcal{C}_{22}) \right) \preceq
    \left( (\mathcal{C}_{11} \otimes \mathcal{C}_{12}) \wedge
        (\mathcal{C}_{21} \otimes \mathcal{C}_{22}) \right)$} \\
\hline
\end{tabular}
\label{tab:meta-props}
\end{table}

\paragraph{Contract refinement.} 

For two contracts $\mathcal{C}_1 = (E_1, M_1)$ and
$\mathcal{C}_2 = (E_2, M_2)$,
$\mathcal{C}_1$ is said to \emph{refine} $\mathcal{C}_2$,
denoted $\mathcal{C}_1 \preceq \mathcal{C}_2$, iff
$M_1 \subseteq M_2$ and $E_2 \subseteq E_1$.
%
As an axiom of the meta-theory, it is required
that the greatest lower bound with respect to
refinement exists, for all subsets of $\mathbb{C}$.
Table~\ref{tab:meta-props} summarises the important
properties of refinement and the other operations on
contracts that a concrete contract theory needs to possess
in order to be considered an instance of the meta-theory. 

\paragraph{Contract conjunction.} 

The \emph{conjunction} of two contracts $\mathcal{C}_1$
and~$\mathcal{C}_2$, 
denoted $\mathcal{C}_1 \wedge \mathcal{C}_2$,
is defined as their greatest lower bound 
w.r.t.\ the refinement order.
(The intention is that $(E_1, M_1) \wedge (E_2, M_2)$
should equal $(E_1 \cup E_2, M_1 \cap M_2)$; however,
this cannot be taken as the definition since not every 
such pair necessarily constitutes a contract.)
Then, we have the three desirable properties of conjunction
listed in Table~\ref{tab:meta-props}, which together are
referred to as \emph{shared refinement}. 

\paragraph{Contract composition.} 

The \emph{composition} of two contracts 
$\mathcal{C}_1 = (E_1, M_1)$
and $\mathcal{C}_2 = (E_2, M_2)$,
denoted 
$\mathcal{C}_1 \otimes \mathcal{C}_2 = (E, M)$, 
is defined when every two components
$m_1 \in M_1$ and~$m_2 \in M_2$ are composable,
and  must then be the least contract,
w.r.t. the refinement order,
satisfying the following conditions: 
\begin{enumerate}[label=(\roman*)]
\item $m_1 \in M_1 \wedge m_2 \in M_2 \Rightarrow m_1 \times m_2 \in M$;
\item $e \in E \wedge m_1 \in M_1 \Rightarrow m_1 \times e \in E_2$; and
\item $e \in E \wedge m_2 \in M_2 \Rightarrow e \times m_2 \in E_2$.
\end{enumerate}
If all of the above is satisfied, then properties~3-6
of Table~\ref{tab:meta-props} hold.
%
The intention is that composing two components
implementing $\mathcal{C}_1$ and~$\mathcal{C}_2$ should
yield an implementation of $\mathcal{C}_1 \otimes \mathcal{C}_2$,
and composing an environment of $\mathcal{C}_1 \otimes \mathcal{C}_2$
with an implementation of $\mathcal{C}_1$ should result in
a valid environment for $\mathcal{C}_2$, and vice versa.
This is important in order to enable independent development.


\section{Denotational Semantics of Programs and Contracts}
\label{sec:preliminaries}

In this section  we summarise the background needed to understand 
the formal developments later in the paper. 
First, we recall the standard denotational semantics of programs
with procedures on a typical toy programming language.
Next, we summarise Hoare logic and contracts,
and provide a semantic justification of procedure-modular verification,
also based on denotational semantics. 


\subsection{The Denotational Semantics of Programs with Procedures}
\label{subsec:denot-sem}

This section sketches the standard presentation
of denotational semantics for procedural languages,
as presented in textbooks such as~\cite{win-93-book,nie-nie-07-book}.
This semantics is the inspiration for the definition
of components in our abstract contract theory in 
Section~\ref{subsec:denot-components}. 
We start with a simplistic programming language not 
involving procedures, and add procedures later to the
language.

The following toy sequential programming language
is typically used to present the denotational semantics
of imperative languages:
$$ S \>::=\> \mathsf{skip} \mid \mathsf{x := a} \mid S_1 ; S_2 \mid
             \mathsf{if}\ b\ \mathsf{then}\ S_1\ \mathsf{else}\ S_2 \mid
             \mathsf{while}\ b\ \mathsf{do}\ S $$
where $S$ ranges over statements, 
$a$~over arithmetic expressions, and
$b$~over Boolean expressions.

To define the denotational semantics of the language, 
we define the set~$\mathbf{State}$ of program states.
A state $s \in \state$ is a mapping from the 
program variables to, for simplicity, the set of integers.

The \emph{denotation} of a statement $S$, denoted $\denot{S}$,
is typically given as a partial function
$\state \hookrightarrow \state$
such that $\denot{S}(s) = s'$ whenever
executing statement~$S$ from the initial state~$s$ 
terminates in state~$s'$.
In case that executing~$S$ from~$s$ does not terminate, 
the value of $\denot{S}(s)$ is undefined. 
The definition of~$\denot{S}$ proceeds by induction on
the structure of~$S$.
For example, the meaning of sequential composition
of statements is usually captured with relation
composition, as given by the equation
$\denot{S_1 ; S_2} \defeq \denot{S_1} \circ \denot{S_2}$. 
%
For the treatment of the remaining statements of the
language, the reader is referred 
to~\cite{win-93-book,nie-nie-07-book}.
%

The definition of denotation captures through its type
(as a partial function) that the execution of statements 
is deterministic. For non-deterministic programs, the
type of denotations is relaxed to 
$\denot{S} \subseteq \cartprodstate$; then,
$(s, s') \in \denot{S}$ captures that there is an
execution of~$S$ starting in~$s$ that terminates in~$s'$.
For technical reasons that will become clear below, we
shall use this latter denotation type in our treatment.

Note that we could alternatively have chosen $\trace$
as the denotational domain, and most results would
still hold in the context of finite-trace semantics.
However, we chose to develop the theory with a focus
on Hoare-logic and deductive verification.
In fact, the domain $\cartprodstate$ can be seen
as a special case of finite traces.
In future work, we will also investigate concrete
contract languages based on this semantics, and
extend the theory for that context.

\paragraph{Procedures and Procedure Calls.}

To extend the language and its denotational semantics
with procedures and procedure calls, we
follow again the approach of~\cite{win-93-book}, but
adapt it to an ``open'' setting, where some called 
procedures might not be declared.
We consider programs in the context of a finite
set~$\mathcal{P}$ of procedure names 
(of some larger, ``closed'' program),
and a set of \emph{procedure declarations}
of the form $\mathsf{proc}\ p\ \mathsf{is}\ S_p$, 
where $p \in \mathcal{P}$.
Further, we extend the toy programming language with the
statement $\mathsf{call}\ p$.
%

\begin{lstlisting}[
    frame=single,
    basicstyle=\small,
    framesep=5pt,
    label={lst:while-even-odd},
    caption={An even-odd toy program.}
]
proc even is if n = 0 then r := 1 else (n := n - 1; call odd);
proc odd  is if n = 0 then r := 0 else (n := n - 1; call even)
\end{lstlisting}
~

As an example, Listing~\ref{lst:while-even-odd} shows a (closed) program in
the toy language, implementing two mutually recursive procedures.
The procedures check whether the value of the global variable~$n$
is even or odd, respectively, and assign the corresponding
truth value to the variable~$r$.
\comment{
The denotation of, say, $\mathit{even}$ will include pairs that
relate states where $n = 0$ to states where all variables
have the same value as in the pre-state, except that $r = 1$.
However, for all other values of~$n$, the denotation
will depend on the call to $\mathit{odd}$.
\todo{Improve this.}
} 

Due to the (potential) recursion in the procedure declarations,
the denotation of $\mathsf{call}\ p$, and thus of the whole 
language, cannot be defined by structural induction as directly 
as before. 
%
%
We therefore define, for any set $P \subseteq \mathcal{P}$
of procedure names, the set $\fenvtype{P}$ of 
\emph{procedure environments},
each environment $\rho \in \fenv{P}$ thus providing a denotation
for each procedure in~$P$.

Let $\env \defeq \bigcup_{P \subseteq \mathcal{P}} \fenv{P}$
be the set of all procedure environments.
We define a partial order relation~$\sqsubseteq$
on procedure environments, as follows.
For any two procedure environments
$\rho \in \fenv{P}$ and $\rho' \in \fenv{P'}$,
$\rho \sqsubseteq \rho'$ if and only if $P \subseteq P'$
and $\forall p \in P.\ \rho(p) \subseteq \rho'(p)$.

Recall that a \emph{complete lattice} is a partial order,
every set of elements of which has a greatest lower bound 
(\emph{glb}) within the domain of the lattice (see, 
e.g.,~\cite{win-93-book}).
It is easy to show that for any $P \subseteq \mathcal{P}$,
$(\fenv{P}, \sqsubseteq)$ is  a complete lattice,
since a greatest lower bound will exist within $\fenv{P}$.
Then, the least upper bound (\emph{lub})
$\rho_1 \sqcup \rho_2$ of any two function environments
$\rho_1 \in \fenv{P_1}$ and $\rho_2 \in \fenv{P_2}$
also exists, and is the environment
$\rho \in \fenv{P_1 \cup P_2}$ such that
$\forall p \in P_1 \cup P_2.\ \rho (p) = \rho_1 (p) \cup \rho_2 (p)$.

We will sometimes need a procedure environment that
maps every procedure in~$P$ to~$\cartprodstate$, and
we shall denote this environment by~$\rho^\top_{P}$.

Next, for sets of procedures, we shall need the notion
of \emph{interface}, which is a pair $\intfpair{}$ of
disjoint sets of procedure names, where
$\fprov{} \subseteq \mathcal{P}$ is a set of \emph{provided}
(or declared) procedures, and
$\freq{} \subseteq \mathcal{P}$ a set of \emph{required}
(or called, but not declared) ones.

Then, we (re)define the notion of denotation of statements~$S$
in the context of a given interface $\intfpair{}$ and 
environments~$\rho^- \in \fenv{\freq{}}$ 
and~$\rho^+ \in \fenv{\fprov{}}$, and denote it
by~$\denot{S}_{\rho^-}^{\rho^+}$. 
In particular, we define
$\denot{\mathsf{call}\ p}_{\rho^-}^{\rho^+}$ as~$\rho^- (p)$
when $p \in P^-$ and as~$\rho^+ (p)$ when $p \in P^+$.

Intuitively, the denotation of a call to a procedure should 
be equal to the denotation of the body of the latter.
\comment{
We therefore form the system of equations:
$$ \set{X_p = \denot{S_p}_{\pi}}_{p \in P} $$
One can view the solutions of this system as environments,
and~$\pi$ as a variable; then, this equation system can be
seen as \emph{one} equation, but over environments. 
} 
We therefore introduce,
given an environment $\rho^- \in \fenv{\freq{}}$,
the function
$\xi : \fenv{\fprov{}} \rightarrow \fenv{\fprov{}}$
defined by
$\xi (\rho^+) (p) \defeq \denot{S_p}_{\rho^-}^{\rho^+}$
for any $\rho^+ \in \fenv{\fprov{}}$ and~$p \in \fprov{}$,
%
%
and consider its fixed points.
By the Knaster-Tarski Fixed-Point Theorem
(as stated, e.g., in~\cite{win-93-book}), 
since $(\fenv{\fprov{}}, \sqsubseteq)$ is a complete lattice
and $\xi$ is monotonic,
$\xi$ has a least fixed-point~$\rho^+_0$.
%

Finally, we define the notion of \emph{standard denotation} 
of statement~$S$ in the context of a given interface $\intfpair{}$ and 
environment~$\rho^- \in \fenv{\freq{}}$, denoted
$\denot{S}_{\rho^-}$, by 
$\denot{S}_{\rho^-} \defeq \denot{S}_{\rho^-}^{\rho^+_0}$,
where $\rho^+_0$ is the least fixed-point defined above. 
%

For example, for the closed program in 
Listing~\ref{lst:while-even-odd}, we have an interface with
$\fprov{} = \{ \mathit{even}, \mathit{odd} \}$
and $\freq{} = \varnothing$.
Then, $(s, s') \in \denot{S_\mathit{even}}_{\rho^-}^{\rho^+}$
if either $s(n)=0$ and $s' = s[r \mapsto 1]$, or else if
$s(n)>0$ and
$(s[n \mapsto s(n)-1], s') \in \rho^+ (\mathit{odd})$.
The denotation $\denot{S_\mathit{odd}}_{\rho^-}^{\rho^+}$ 
is analogous.
The resulting least fixed-point~$\rho^+_0$ is such that
$(s, s') \in \denot{S_\mathit{even}}_{\rho^-}$,
or equivalently 
$(s, s') \in \denot{S_\mathit{even}}_{\rho^-}^{\rho^+_0}$,
whenever $s(n) \geq 0$, and either $s(n)$ is even and
then $s'(n) = 0$ and  $s'(r) = 1$, or else $s(n)$ is
odd and then $s'(n) = 0$ and  $s'(r) = 0$.
The standard denotation $\denot{S_\mathit{odd}}_{\rho^-}$
of $\mathit{odd}$ is analogous.


\subsection{Hoare Logic and Contracts}
\label{subsec:prog-contr}

In this section we summarise the denotational semantics of
Hoare logic and the semantic
justification of procedure-modular verification, as developed
by the second author in~\cite{gurov-18}.
These formalisations serve as the starting
point for the definition of contracts in our contract theory
developed in Section~\ref{subsec:denot-contracts}.

\paragraph{Hoare Logic.}

The basic judgement of
Hoare logic~\cite{hoare-69} is the Hoare triple,
written $\{P\} S \{Q\}$, where $P$ and~$Q$ are
assertions over the program state, and~$S$ is a
program statement. The Hoare triple signifies that
if the statement~$S$ is executed from a state that
satisfies~$P$ (called the pre-condition), and if
this execution terminates, then the final state
of the execution will satisfy~$Q$ (called the
post-condition).
Additionally, so-called \emph{logical variables}
can be used within a Hoare triple, to specify the
desired relationship between the values of variables
after execution and the values of variables before
execution.
The values of the program variables are defined
by the notion of state; to give a meaning to the
logical variables we shall use 
\emph{interpretations}~$\mathcal{I}$.
We shall write $s \models_\mathcal{I} P$ to signify
that the assertion~$P$ is true w.r.t.\ state~$s$
and interpretation~$\mathcal{I}$.
The formal validity of a Hoare triple is denoted
by $\models_{\mathit{par}} \{P\} S \{Q\}$, where
the subscript signifies that validity is in terms
of \emph{partial correctness}, where termination
of the execution of~$S$ is not required.

An example of a Hoare triple, stating the desired
behaviour of procedure~$\mathit{odd}$ from
Listing~\ref{lst:while-even-odd}, is shown below,
where we use the logical variable~$n_0$ to capture
to the value of~$n$ prior to execution of~$\mathit{odd}$:
\begin{equation}
\label{eq:odd-triple}
\{ n \geq 0 \wedge n = n_0 \}\ S_\mathit{odd}\ \{ 
    (n_0\ \mathrm{mod}\ 2 = 0 \Rightarrow r = 0) \wedge
    (n_0\ \mathrm{mod}\ 2 = 1 \Rightarrow r = 1)
\}
\end{equation}
Procedure~$\mathit{even}$ is specified analogously.

Hoare logic comes with a proof calculus
for reasoning in terms of Hoare triples,
consisting of proof rules for the different
types of statements of the programming language.
An example is the rule for sequential composition:
$$
\infer[\ruleName{Composition}]
{\hoare{P}{S_1; S_2}{Q}}
{\hoare{P}{S_1}{R} &\quad \hoare{R}{S_2}{Q}}
$$
which essentially states that if executing~$S_1$
from any state satisfying~$P$ terminates (if at all) 
in some state satisfying~$R$, and executing~$S_2$ from
any state satisfying~$R$ terminates (if at all) in 
some state satisfying~$Q$, then it is the case that 
executing the composition $S_1; S_2$ from any state
satisfying~$P$ terminates (if at all) in some state 
satisfying~$Q$. 
The proof system is sound and relatively complete w.r.t.\
the denotational semantics of the programming 
language (see, e.g.,~\cite{win-93-book,nie-nie-07-book}). 

\paragraph{Hoare Logic Contracts.}

One can view a Hoare triple $\{P\} S \{Q\}$
as a \emph{contract} $C = (P, Q)$ imposed on the program~$S$.
In many contexts it is meaningful to separate the
contract from the program; for instance, if the
program is yet to be implemented. 
In our earlier work~\cite{gurov-18}, we gave such contracts 
a denotational semantics as follows:
\begin{equation}
\label{eq:hoare-contract}
\denot{C} \defeq \setdef{(s, s')}{\forall \mathcal{I}.\ 
   (s \models_\mathcal{I} P \Rightarrow s' \models_\mathcal{I} Q)}
\end{equation}
The rationale behind this definition is the
following desirable property:
a program \emph{meets} a contract whenever its denotation is
subsumed by the denotation of the contract, i.e.,
$ S \models_{\mathit{par}} C $ 
if and only if
$ \denot{S} \subseteq \denot{C} $.
   
For example, for the contract~$C_\mathit{odd}$ induced
by~(\ref{eq:odd-triple}) we have that
$(s, s') \in \denot{C_\mathit{odd}}$ if and only if
either $s(n) < 0$, or else $s' (r) = 0$ if
$s (n)$ is even and $s' (r) = 1$ if $s (n)$ is odd.
The denotation of~$C_\mathit{even}$ is analogous.

\paragraph{The Denotational Semantics of Programs with
Procedure Contracts.}

Let~$S$ be a program with procedures, and let
every declared procedure~$p \in \mathcal{P}$ be equipped
with a procedure contract~$C_p$.
\emph{Procedure-modular verification} refers to
techniques that verify every procedure in isolation.
The key to this is to handle procedure calls by
using the contract of the called procedure
rather than its body (i.e., by \emph{contracting}
rather than by \emph{inlining}~\cite{bub-et-al-14}).
In~\cite{gurov-18},
a semantic justification of this is given by means of a
\emph{contract-relative} denotational semantics of
statements. The intuition behind this semantics is
that procedure calls are given a meaning through
the denotation of the contract of the called
procedure, rather than through the denotation of
its body.

The contract-relative denotational semantics of a 
statement~$S$, 
denoted $\denot{S}^{cr}$, is defined with the
help of the \emph{contract environment}~$\rho_c$
that is induced by the procedure contracts, i.e.,
$\rho_c (p) \defeq \denot{C_p}$ for all
$p \in \mathcal{P}$, as 
$\denot{S}^{cr} \defeq \denot{S}_{\rho_c}$.
Notice that this definition does not involve
solving any recursive equations (i.e., finding 
fixed points),
and gives rise to a contract-relative notion of when
a statement meets a contract, namely 
$S \models_{\mathit{par}}^{\mathit{cr}} C$
if and only if
$\denot{S}^{cr} \subseteq \denot{C}$.
This is exactly the correctness notion that
is the target of procedure-modular verification.
As shown in~\cite{gurov-18}, this notion is \emph{sound}
w.r.t.\ the original notion $S \models_{\mathit{par}} C$,
in the sense that $ S \models_{\mathit{par}}^{\mathit{cr}} C$
entails $S \models_{\mathit{par}} C$. In other words,
verifying a program procedure-modularly establishes
that the program is correct w.r.t.\ its contract in
the standard sense.

For example, the contract-relative semantics of~$S_\mathit{even}$
is such that $(s, s') \in \denot{S_\mathit{even}}^{cr}$
if either $s(n) < 0$, or
$s(n) = 0$ and $s' = s [r \mapsto 1]$, or else
$s' (r) = 1$ if $s (n)$ is even and
$s' (r) = 0$ if $s (n)$ is odd.
The contract-relative semantics of~$S_\mathit{odd}$ is
analogous.
Then, it is easy to check that both
$S_\mathit{even} \models_{\mathit{par}}^{\mathit{cr}} C_\mathit{even}$
and
$S_\mathit{odd} \models_{\mathit{par}}^{\mathit{cr}} C_\mathit{odd}$
hold.

\comment{
using the triple given in~(\ref{eq:odd-triple})
and the definition in~(\ref{eq:hoare-contract}),
the denotation of the contract of \texttt{odd} will
map states where $n$ is odd to states
where $r = 1$ (and $r = 0$ when $n$ is even),
and under those restrictions it will contain all
combinations of changes to valuations of other
variables, since they are unconstrained in the Hoare triple.
Given that, the contract-relative denotation of
\texttt{even} will be different to (in fact, a
strict superset of) the environment
relative denotation from Section~\ref{subsec:denot-sem},
in that it will also contain all combinations of
valuations of unaffected variables (except in the
case $n = 0$), since according to the contract of
\texttt{odd} they might change during the call.
However, if the contract of \texttt{even} is specified
similarly to \texttt{odd}'s, then both denotations
will be a subset of the contract denotation.
} 



\section{An Abstract Contract Theory}
\label{sec:proc-lang-contract-theory}

This section presents an abstract contract theory for
programs with procedures.
The theory builds on the basic notion of
\emph{denotation} as a binary relation over states.
As we will show later, it is
both an abstraction of the denotational semantic view
on programs with procedures and procedure contracts
presented in Sections~\ref{subsec:denot-sem}
and~\ref{subsec:prog-contr}, and an instantiation of the
meta-theory described in Section~\ref{sec:meta-theory}.



\subsection{Components}
\label{subsec:denot-components}

In the context of a concrete programming language,
we view a component as a module, consisting of a
collection of procedures that are \emph{provided} by
the module. The module may call \emph{required}
procedures that are external to the module. The
way the provided procedures transform the program 
state upon a call depends on how the required 
procedures transform the state. 
We take this observation as the basis of our
abstract setting, in which state transformers are
modelled as denotations (i.e., as binary relations
over states).
%
A component will thus be simply a mapping from
denotations of the required procedures to denotations
of the provided ones, both captured through the notion
of procedure environments.

The contract theory is abstract, in that it is not
defined for a particular programming language, and
may be instantiated with any procedural language.
As with the meta-theory, the abstract contract theory
is also defined only on the semantic level.

Recall the notions and notation from 
Section~\ref{subsec:denot-sem}.
A component \emph{interface} $\intfeq{}$
is a  pair of disjoint, finite sets of procedure names,
of the required and the provided ones, respectively.

\begin{definition}[Component]
\label{def:component}
A \emph{component}~$m$ with interface 
$\intfeq{m}$ is a mapping $\comptype{m}$. 
\end{definition}
\noindent
Let~$\mathcal{M}$ denote the universe of all components
over~$\mathcal{P}$.

We assume that any system is built up from a set
of \emph{base components}, the simplest components
from which more complex components are then obtained
by composition.
The base components must be \emph{monotonic} functions
over the lattice defined in Section~\ref{subsec:denot-sem}.

When $\freq{m} = \varnothing$, we shall identify~$m$ 
with an element of $\fenv{\fprov{m}}$.
In other words, when a component is \emph{closed}, i.e., is not dependent
on any external procedures, the provided environment
is constant.

\begin{definition}[Component composability]
\label{def:comp-composability}
Two components $m_1$ and $m_2$ are \emph{composable}
iff $\fprov{m_1} \cap \fprov{m_2} = \varnothing$.
\end{definition}

When defining the composition of two components, particular
care is required in the treatment of procedure names that
are provided by one of the components while required by 
the other. 
Let $\mu x.\ f(x)$
denote the least fixed-point of a 
function~$f$, when it exists.

\begin{definition}[Component composition]
\label{def:comp-composition}
Given two composable components
$\comptype{m_1}$ and $\comptype{m_2}$,
their \emph{composition} is defined as a mapping
$\comptype{m_1 \times m_2}$ such that:
\begin{align*}
\fprov{m_1 \times m_2} \defeq \ & \fprov{m_1} \cup \fprov{m_2} \\
\freq{m_1 \times m_2} \defeq \ &
(\freq{m_1} \cup \freq{m_2}) \setminus (\fprov{m_1} \cup \fprov{m_2}) \\
m_1 \times m_2 \defeq \ &
\lambda \anyreq{\rho}{m_1 \times m_2} \in \fenv{\freq{m_1 \times m_2}}.
\ \mu \rho.\ \anyprov{\chi}{m_1 \times m_2}(\rho)
\end{align*}
where 
$\anyprov{\chi}{m_1 \times m_2} : \fenv{\fprov{m_1 \times m_2}}
 \rightarrow \fenv{\fprov{m_1 \times m_2}}$
is defined, in the context of a given
$\anyreq{\rho}{m_1 \times m_2} \in \fenv{\freq{m_1 \times m_2}}$,
as follows. Let
$\anyprov{\rho}{m_1 \times m_2} \in \fenv{\fprov{m_1 \times m_2}}$,
and let $\anyreq{\rho}{m_1} \in \fenv{\freq{m_1}}$
be the environment defined by:
$$ \anyreq{\rho}{m_1} (p) \defeq
     \left\{ 
     \begin{array}{ll}
     \anyprov{\rho}{m_1 \times m_2} (p)~~~ & \mbox{if $p \in \freq{m_1} \cap \fprov{m_2}$} \\
     \anyreq{\rho}{m_1 \times m_2} (p) & \mbox{if $p \in \freq{m_1} \setminus \fprov{m_2}$} \\
     \end{array}
     \right. $$
and let $\anyreq{\rho}{m_2} \in \fenv{\freq{m_2}}$
be defined symmetrically. We then define:
$$ \anyprov{\chi}{m_1 \times m_2} (\anyprov{\rho}{m_1 \times m_2}) (p) 
   \defeq
     \left\{ 
     \begin{array}{ll}
     m_1 (\anyreq{\rho}{m_1}) (p)~~~ & \mbox{if $p \in \fprov{m_1}$} \\
     m_2 (\anyreq{\rho}{m_2}) (p)~~~ & \mbox{if $p \in \fprov{m_2}$} \\
     \end{array}
     \right. $$
\end{definition}

In the above definition, $\anyprov{\chi}{m_1 \times m_2}$
represents the denotations of the procedure \emph{bodies}
of the procedures provided by the two composed components,
given denotations of procedure \emph{calls} to the same
procedures.
The choice of least fixed-point will be crucial for the
proof of 
Theorem~\ref{thm:contract-composition}\ref{itm:denot-comp-impl}
in Section~\ref{subsec:denot-contracts} below. 


The definition is well-defined, in the sense that
the stated least fixed-points exist, and the resulting
components are monotonic functions. 

\begin{theorem}
\label{thm:comp-welldef}
Component composition is well-defined.
\end{theorem}

\begin{proof}
%
For any $P \subseteq \mathcal{P}$, 
we have that $(\fenv{P}, \sqsubseteq)$ forms
a complete lattice.
Thus, in particular,
$(\fenv{\fprov{m_1 \times m_2}}, \sqsubseteq)$
is a complete lattice.
By the Knaster-Tarski Fixed-Point Theorem,
as stated e.g.\ in~\cite{win-93-book}, 
if $(\fenv{\fprov{m_1 \times m_2}}, \sqsubseteq)$
is a complete lattice, and
$\anyprov{\chi}{m_1 \times m_2} :
\fenv{\fprov{m_1 \times m_2}} \rightarrow
\fenv{\fprov{m_1 \times m_2}}$
is monotonic, then $\anyprov{\chi}{m_1 \times m_2}$
has a least fixed-point.

To prove that components are monotonic functions,
we shall use structural induction. Since base components
are monotonic by definition, we only have to show that
composition preserves monotonicity. 
Assume that $m_1$ and $m_2$ are monotonic.
Since $\anyprov{\chi}{m_1 \times m_2}$ is an
application of either $m_1$ or $m_2$,
it must also be monotonic, and thus have a least fixed-point. 
The $\mu$ operator is itself monotonic,
and the component $m_1 \times m_2$ must therefore be monotonic as well.
\end{proof}

\subsection{Denotational Contracts}
\label{subsec:denot-contracts}

We now define the notion of denotational contracts~$c$
in the style of \emph{assume/guarantee
contracts}~\cite{ben-et-al-07,benvenuti-et-al-08}.
Contracts shall also be given interfaces.

\begin{definition}[Denotational contract]
\label{def:denot-contract}
A \emph{denotational contract}~$c$ with interface 
$\intfeq{c}$ is a pair $(\rho_c^-, \rho_c^+)$, where 
$\rho_c^- \in \fenv{\freq{c}}$ and $\rho_c^+ \in \fenv{\fprov{c}}$.
\end{definition}
\noindent
The intended interpretation of the environment 
pair is as follows:
\emph{assuming} that the denotation of every
called procedure~$p \in \freq{c}$ 
is subsumed by $\rho_c^- (p)$, then 
it is \emph{guaranteed} that the denotation of every
provided procedure~$p' \in \fprov{c}$ 
is subsumed by $\rho_c^+ (p')$. 

\begin{definition}[Contract implementation]
\label{def:denot-contract-imp}
A component~$m$ with interface $\intfeq{m}$ 
is an \emph{implementation} for,
or \emph{implements},
a contract~$c = (\rho_c^-, \rho_c^+)$
with interface $\intfeq{c}$,
denoted $m \models c$, iff
$\freq{c} \subseteq \freq{m}$, 
$\fprov{m} \subseteq \fprov{c}$,
and $m(\rho_c^- \sqcup \rho^\top_{\freq{m} \setminus \freq{c}})
    \sqsubseteq \rho_c^+$.
\end{definition}
\noindent
The reason for not requiring the interfaces to be equal
is that we aim at a subset relation between
components implementing a contract and those implementing
a refinement of said contract, in the meta-theory instantiation.


For a mapping $h : A \rightarrow B$ and set $A' \subseteq A$, 
let $h_{|A'}$ denote as usual the restriction of~$h$ on~$A'$.

\begin{definition}[Contract environment]
\label{def:denot-contract-env}
A component~$m$ is an \emph{environment} for
contract~$c$ iff, for any implementation~$m'$ of $c$,
$m$ and $m'$ are composable, and
$\forall \anyreq{\rho}{m \times m'} \in \fenv{\freq{m \times m'}}.\ 
(m \times m')(\anyreq{\rho}{m \times m'})_{|\fprov{c}}
\sqsubseteq \anyprov{\rho}{c}$.
\end{definition}
\noindent
Intuitively, an environment of a contract~$c$ is
then a component such that when it is composed with
an implementation of~$c$, the composition
will operate satisfactorily with respect to the
guarantee of the contract.
%
%

We will now define the refinement relation,
and the conjunction and composition operations,
on contracts.

\begin{definition}[Contract refinement]
\label{def:contract-refinement}
A contract $c$ \emph{refines} contract $c'$,
denoted $c \preceq c'$,
iff $\rho_{c'}^- \sqsubseteq \rho_{c}^-$ and
$\rho_{c}^+ \sqsubseteq \rho_{c'}^+$, where
$\sqsubseteq$ is the partial order relation
defined in Section~\ref{subsec:denot-sem}.
\end{definition}
\noindent
The refinement relation reflects the intention that if
a contract~$c$ refines another contract~$c'$, then
any component implementing~$c$ should also implement~$c'$.

\begin{definition}[Contract conjunction]
\label{def:contract-conjunction}
The \emph{conjunction} of two contracts
$c_1 = (\rho_{c_1}^-, \rho_{c_1}^+)$ 
and 
$c_2 = (\rho_{c_2}^-, \rho_{c_2}^+)$
is the contract
$c_1 \wedge c_2 \defeq (\rho_{c_1}^- \sqcup \rho_{c_2}^-, \rho_{c_1}^+ \sqcap \rho_{c_2}^+)$,
where~$\sqcup$ and~$\sqcap$ are the \emph{lub} and 
\emph{glb} operations of the lattice, respectively. 
\end{definition}
\noindent
This definition is consistent with the intention that 
any contract that refines $c_1 \wedge c_2$ should also 
refine~$c_1$ and~$c_2$ individually.
The interface of $c_1 \wedge c_2$ is then
$\intfsym{c_1 \wedge c_2} =
(\freq{c_1} \cup \freq{c_2}, \fprov{c_1} \cap \fprov{c_2})$.
Note that while this is the interface in general,
conjunction of contracts is typically used to
merge different viewpoints of \emph{the same} component,
and in that case
$\intfsym{c_1} = \intfsym{c_2} = \intfsym{c_1 \wedge c_2}$. 

\begin{definition}[Contract composability]
\label{def:contract-composability}
Two contracts
$c_1 = (\anyreq{\rho}{c_1}, \anyprov{\rho}{c_1})$ and
$c_2 = (\anyreq{\rho}{c_2}, \anyprov{\rho}{c_2})$
with interfaces $\intfeq{c_1}$ and $\intfeq{c_2}$
are \emph{composable} if:
\begin{enumerate*}[label=(\roman*)]
\item $\fprov{c_1} \cap \fprov{c_2} = \varnothing$,
\item $\forall p \in \freq{c_1} \cap \fprov{c_2}.\ 
       \rho_{c_2}^+(p) \subseteq \rho_{c_1}^-(p)$, and\linebreak
\item $\forall p \in \freq{c_2} \cap \fprov{c_1}.\
       \rho_{c_1}^+(p) \subseteq \rho_{c_2}^-(p)$.
\end{enumerate*}
\end{definition}
\noindent
The conditions for composability ensure that 
the mutual guarantees of the two contracts 
meet each other's assumptions.

\begin{definition}[Contract composition]
\label{def:contract-composition}
The \emph{composition} of two composable contracts
$c_1 = (\rho_{c_1}^-, \rho_{c_1}^+)$ 
and 
$c_2 = (\rho_{c_2}^-, \rho_{c_2}^+)$,
with interfaces 
$\intfeq{c_1}$ and $\intfeq{c_2}$, respectively,
is the contract
$c_1 \otimes c_2 \defeq (\rho_{c_1 \otimes c_2}^-, \rho_{c_1}^+ \sqcup \rho_{c_2}^+)$, where:
\begin{equation*}
    \anyreq{\rho}{c_1 \otimes c_2} \defeq
        (\anyreq{\rho}{c_1} \sqcap \anyreq{\rho}{c_2})_{\big|
           (\freq{c_1} \cup \freq{c_2})
            \setminus (\fprov{c_1} \cup \fprov{c_2})}
\end{equation*}
\end{definition}
\noindent
The interface of $c_1 \otimes c_2$ is $\intfsym{c_1 \otimes c_2} =
((\freq{c_1} \cup \freq{c_2}) \setminus (\fprov{c_1} \cup \fprov{c_2}),
\fprov{c_1} \cup \fprov{c_2})$.
\begin{theorem}
\label{thm:contract-composition}
For any composable contracts $c_1$ and~$c_2$, and any
implementations $m_1 \models c_1$ and $m_2 \models c_2$,
$m_1$ and~$m_2$ are composable, and $c_1 \otimes c_2$ is
the least contract (w.r.t. refinement order) for which
the following properties hold:
\begin{enumerate}[label=(\roman*)]
    \item \label{itm:denot-comp-impl}
        $m_1 \times m_2 \models c_1 \otimes c_2$,
    \item \label{itm:denot-comp-env1}
        if $m$ is an environment to $c_1 \otimes c_2$,
        then $m_1 \times m$ is an environment to $c_2$,
    \item \label{itm:denot-comp-env2}
        if $m$ is an environment to $c_1 \otimes c_2$,
        then $m \times m_2$ is an environment to $c_1$.
\end{enumerate}
\end{theorem}

\begin{proof}
Let the contracts
$c_1 = (\rho_{c_1}^-, \rho_{c_1}^+)$ 
and 
$c_2 = (\rho_{c_2}^-, \rho_{c_2}^+)$
with interfaces 
$\intfeq{c_1}$ and $\intfeq{c_2}$
be composable, and let 
$m_1 \models c_1$ and $m_2 \models c_2$. 
Since contract composition restricts the interface in the
same way as component composition, $m_1$ and~$m_2$ are 
also composable.

Proof of~\ref{itm:denot-comp-impl}. 
Since $m_1 \models c_1$ and $m_2 \models c_2$,
it must be the case that
$\freq{c_1} \cup \freq{c_2} \subseteq \freq{m_1} \cup \freq{m_2}$ and
$\fprov{m_1} \cup \fprov{m_2} \subseteq \fprov{c_1} \cup \fprov{c_2}$,
so
$\freq{c_1 \otimes c_2} \subseteq \freq{m_1 \times m_2}$ and
$\fprov{m_1 \times m_2} \subseteq \fprov{c_1 \otimes c_2}$,
which are the first two conditions from 
Definition~\ref{def:denot-contract-imp}.
Now, let $\rho_{m_1 \times m_2}^- \defeq
\rho_{c_1 \otimes c_2}^- \sqcup
\rho^\top_{\freq{m_1 \times m_2} \setminus \freq{c_1 \otimes c_2}}$
and $\anyprov{\rho}{c_1 \otimes c_2} \defeq \rho_{c_1}^+ \sqcup \rho_{c_2}^+$.
Recall the definition of
$\anyprov{\chi}{m_1 \times m_2}$ in 
Definition~\ref{def:comp-composition}, 
defined now in the context of the above 
$\rho_{m_1 \times m_2}^-$.
We need to show
$(m_1 \times m_2) (\anyreq{\rho}{m_1 \times m_2})
\sqsubseteq \anyprov{\rho}{c_1 \otimes c_2}$, i.e., that
$\mu \rho.\ \anyprov{\chi}{m_1 \times m_2}(\rho)
\sqsubseteq \anyprov{\rho}{c_1 \otimes c_2}$.
By using that 
$\anyprov{\rho}{c_1 \otimes c_2|\fprov{m_1 \times m_2}}
\sqsubseteq \anyprov{\rho}{c_1 \otimes c_2}$,
this reduces to showing 
$\mu \rho.\ \anyprov{\chi}{m_1 \times m_2}(\rho)
\sqsubseteq \anyprov{\rho}{c_1 \otimes c_2|\fprov{m_1 \times m_2}}$.
This we accomplish by proving that 
$\anyprov{\chi}{m_1 \times m_2}
(\anyprov{\rho}{c_1 \otimes c_2|\fprov{m_1 \times m_2}})
\sqsubseteq \anyprov{\rho}{c_1 \otimes c_2|\fprov{m_1 \times m_2}}$,
which establishes that
$\anyprov{\rho}{c_1 \otimes c_2|\fprov{m_1 \times m_2}}$
is a prefixed-point of $\anyprov{\chi}{m_1 \times m_2}$,
and is therefore greater than 
$\mu \rho.\ \anyprov{\chi}{m_1 \times m_2}(\rho)$.

Thus, we need to show
$\anyprov{\chi}{m_1 \times m_2}
(\anyprov{\rho}{c_1 \otimes c_2|\fprov{m_1 \times m_2}})
\sqsubseteq \anyprov{\rho}{c_1 \otimes c_2|\fprov{m_1 \times m_2}}$,
i.e., that
$\forall p \in \fprov{m_1 \times m_2}.\ \anyprov{\chi}{m_1 \times m_2}
(\anyprov{\rho}{c_1 \otimes c_2|\fprov{m_1 \times m_2}}) (p)
\subseteq \anyprov{\rho}{c_1 \otimes c_2|\fprov{m_1 \times m_2}} (p)$.
%
%
Let~$p \in \fprov{m_1}$; the case~$p \in \fprov{m_2}$ 
is symmetric. 
Using the notation of Definition~\ref{def:comp-composition}, 
we have:
$$ \anyreq{\rho}{m_1} (p') =
     \left\{ 
     \begin{array}{ll}
     \anyprov{\rho}{c_1 \otimes c_2|\fprov{m_1 \times m_2}} (p')~~~ & \mbox{if $p' \in \freq{m_1} \cap \fprov{m_2}$} \\
     (\rho_{c_1 \otimes c_2}^- \sqcup \rho^\top_{\freq{m_1 \times m_2} 
     \setminus \freq{c_1 \otimes c_2}}) (p')~~~ & \mbox{if $p' \in \freq{m_1} \setminus \fprov{m_2}$} \\
     \end{array}
     \right. $$
Below we shall show 
$\anyreq{\rho}{m_1} \sqsubseteq \anyreq{\rho}{c_1} \sqcup \rho^\top_{\freq{m_1} \setminus \freq{c_1}}$. Since
$m_1 \models c_1$, by Definition~\ref{def:denot-contract-imp}
we have 
$m(\rho_{c_1}^- \sqcup \rho^\top_{\freq{m_1} \setminus \freq{c_1}})
    \sqsubseteq \rho_{c_1}^+$, and by monotonicity of~$m_1$
and transitivity of~$\sqsubseteq$, we obtain
$m(\anyreq{\rho}{m_1}) \sqsubseteq \anyprov{\rho}{c_1}$. 
Then, for~$p \in \fprov{m_1}$, we have
$\anyprov{\chi}{m_1 \times m_2}
(\anyprov{\rho}{c_1 \otimes c_2|\fprov{m_1 \times m_2}}) (p) = m_1 (\anyreq{\rho}{m_1}) (p) \subseteq (\anyprov{\rho}{c_1}) (p) = \anyprov{\rho}{c_1 \otimes c_2|\fprov{m_1 \times m_2}} (p)$, which
is what we wanted to prove. 
    
And finally, we need to show
$\anyreq{\rho}{m_1} \sqsubseteq \anyreq{\rho}{c_1}
\sqcup \rho^\top_{\freq{m_1} \setminus \freq{c_1}}$.
There are altogether five cases that need to be considered.
First, let~$p' \in \freq{c_1} \cap \fprov{m_2}$
(recall that $\freq{c_1} \subseteq \freq{m_1}$, since
$m_1 \models c_1$). 
Since $c_1$ and~$c_2$ are composable, by 
Definition~\ref{def:contract-composability} we have 
$\forall p \in \freq{c_1} \cap \fprov{c_2}.\ 
       \rho_{c_2}^+(p) \subseteq \rho_{c_1}^-(p)$.
Then we have
$\anyreq{\rho}{m_1} (p') = \anyprov{\rho}{c_1 \otimes c_2|\fprov{m_1 \times m_2}} (p') = \rho_{c_2}^+ (p') \subseteq \rho_{c_1}^- (p') = (\anyreq{\rho}{c_1} \sqcup \rho^\top_{\freq{m_1} \setminus \freq{c_1}}) (p')$. 
Next, let~$p' \in (\freq{m_1} \setminus \freq{c_1}) \cap \fprov{m_2}$. 
This case is similar to the first one, but now
$\rho_{c_2}^+ (p') \subseteq \rho^\top_{\freq{m_1} \setminus \freq{c_1}} (p') = (\anyreq{\rho}{c_1} \sqcup \rho^\top_{\freq{m_1} \setminus \freq{c_1}}) (p')$. 
Third, let~$p' \in (\freq{c_1} \setminus \fprov{m_2}) \setminus \freq{c_2}$. 
Then we have
$\anyreq{\rho}{m_1} (p') = (\rho_{c_1 \otimes c_2}^- \sqcup \rho^\top_{\freq{m_1 \times m_2} \setminus \freq{c_1 \otimes c_2}}) (p') = \rho_{c_1}^- (p') = (\anyreq{\rho}{c_1} \sqcup \rho^\top_{\freq{m_1} \setminus \freq{c_1}}) (p')$. 
Fourth, let~$p' \in (\freq{m_1} \setminus \fprov{m_2}) \setminus \freq{c_1}$. 
This case is similar to the third one, but now
$(\rho_{c_1 \otimes c_2}^- \sqcup \rho^\top_{\freq{m_1 \times m_2} \setminus \freq{c_1 \otimes c_2}}) (p') \subseteq \rho^\top_{\freq{m_1} \setminus \freq{c_1}} (p') = (\anyreq{\rho}{c_1} \sqcup \rho^\top_{\freq{m_1} \setminus \freq{c_1}}) (p')$. 
And last, let~$p' \in \freq{c_1} \cap \freq{c_2}$. 
This case is also similar to the third one, but now,
by Definition~\ref{def:contract-composition}, we have
$(\rho_{c_1 \otimes c_2}^- \sqcup \rho^\top_{\freq{m_1 \times m_2} \setminus \freq{c_1 \otimes c_2}}) (p') = (\anyreq{\rho}{c_1} \sqcap \anyreq{\rho}{c_2}) (p') \subseteq \anyreq{\rho}{c_1} (p') = (\anyreq{\rho}{c_1} \sqcup \rho^\top_{\freq{m_1} \setminus \freq{c_1}}) (p')$. 
This concludes the proof of~\ref{itm:denot-comp-impl}. 

Proof of~\ref{itm:denot-comp-env1}. 
Let $m$ be an environment to $c_1 \otimes c_2$.
Then, $\fprov{m} \cap \fprov{c_1 \otimes c_2} = \varnothing$
and consequently $\fprov{m} \cap \fprov{c_1} = \varnothing$,
so $m$ and $m_1$ are composable.
Furthermore, since $\fprov{m} \cap \fprov{c_2} = \varnothing$
and $\fprov{m_1} \cap \fprov{c_2} = \varnothing$ we have
$\fprov{m_1 \times m} \cap \fprov{c_2} = \varnothing$,
thus the condition on the interfaces hold.
Since $\anyprov{\rho}{c_1 \times c_2}$ is the lub of
$\anyreq{\rho}{c_1}$ and $\anyreq{\rho}{c_2}$, and
$\fprov{c_1} \cap \fprov{c_2} = \varnothing$, then denotations
for all $p \in \fprov{c_2}$ are the same in $\anyprov{\rho}{c_2}$
and $\anyprov{\rho}{c_1 \otimes c_2}$.
Thus, we have that
$\forall \anyreq{\rho}{m_1 \times m \times m_2}
\in \fenv{\freq{m_1 \times m \times m_2}}.\ 
\forall p \in \fprov{c_2}.\ (m_1 \times m \times m_2)
(\anyreq{\rho}{m_1 \times m \times m_2})(p)
\subseteq \anyprov{\rho}{c_2}(p)$
must hold, which it does since
$\anyprov{\rho}{c_2} =
\anyprov{\rho}{c_1 \otimes c_2 \big| \fprov{c_2}}$
and $m$ is an environment to $c_1 \otimes c_2$.
So $m_1 \times m$ is an environment to~$c_2$. 

The proof of~\ref{itm:denot-comp-env2} is analogous. 

Finally, $c_1 \otimes c_2$ is the least contract
for which the above properties hold, since weakening
$\anyreq{\rho}{c_1 \otimes c_2}$ or strengthening
$\anyprov{\rho}{c_1 \otimes c_2}$ would immediately
falsify~\ref{itm:denot-comp-impl}.
\end{proof}

%


\section{Connection to Meta-Theory}
\label{sec:meta-connection}

In this section we show that
the abstract contract theory presented
in Section~\ref{sec:proc-lang-contract-theory}
instantiates the meta-theory described in
Section~\ref{sec:meta-theory}. 

In our instantiation of the meta-theory, we consider
as the abstract component universe~$\mathbb{M}$
the same universe of components~$\mathcal{M}$ as defined
in Section~\ref{subsec:denot-components}.
To distinguish the contracts of the meta-theory from
those of the abstract theory, we shall always denote
the former by~$\mathcal{C}$ and the latter by~$c$.  
%
Recall that a contract~$\mathcal{C}$ is a pair
$(E, M)$, where $E, M \subseteq \mathcal{M}$.
The formal connection between the two notions
is established with the following definition. 

\begin{definition}[Induced contract]
\label{def:meta-contract}
Let $c$ be a denotational contract. It
\emph{induces} the contract
$\mcont{c} = \mpair{c}$, where
$\menv{c} \defeq \{ m \in \mathcal{M}\ |\
m \text{ is an environment for } c\}$ 

\noindent
and
$\mimp{c} \defeq \{ m \in \mathcal{M}\ |\ m \models c \}$.
%
\end{definition}
\noindent
Since contract implementation requires that the implementing component's 
provided functions are a subset of the contract's provided functions, 
every component~$m$ such that $\fprov{m} \cap \fprov{c} = \varnothing$
is composable with every component in~$M_c$. 

The definitions of implementation, refinement and conjunction
of denotational contracts make this straightforward
definition of induced contracts possible, so that it
directly results in refinement as set membership and
conjunction as lub w.r.t.\ the refinement order.

\begin{theorem}
\label{thm:meta-theory-inst}
The contract theory of
Section~\ref{sec:proc-lang-contract-theory}
instantiates the meta-theory
of Benveniste et al.~\cite{ben-et-al-18}, in the sense that
composition of components is associative and commutative, and
for any two contracts $c_1$ and $c_2$:
\begin{enumerate}[label=(\roman*)]
    \item \label{itm:meta-inst-ref}
        $c_1 \preceq c_2$ iff $\mcont{c_1}$
        refines $\mcont{c_2}$
        according to the definition of the meta-theory,
    \item \label{itm:meta-inst-conj}
        $\mcont{c_1 \wedge c_2}$ is the conjunction of
        $\mathcal{C}_{c_1}$ and $\mathcal{C}_{c_2}$
        as defined in the meta-theory, and
    \item \label{itm:meta-inst-comp}
        $\mcont{c_1 \otimes c_2}$ is the composition of
        $\mathcal{C}_{c_1}$ and $\mathcal{C}_{c_2}$
        as defined in the meta-theory.
\end{enumerate}
\end{theorem}

\begin{proof}
Statement~\ref{itm:meta-inst-ref} holds since $c_1 \preceq c_2$ means that for
all components $m$, if $m \models c_1$ then $m \models c_2$.
Equivalently, if $m \models^M \mathcal{C}_{c_1}$ then
$m \models^M \mathcal{C}_{c_2}$, which means that
$M_{c_1} \subseteq M_{c_2}$.
Now, $c_1 \preceq c_2$ also means that
all components $m'$ composable with components in $M_{c_2}$
are also composable with all components in $M_{c_1}$,
and thus $E_{c_2} \subseteq E_{c_1}$.
So $c_1 \preceq c_2$ iff
$\mathcal{C}_{c_1} \preceq \mathcal{C}_{c_2}$.

Statement~\ref{itm:meta-inst-conj} clearly holds: any component
implementing $c_1 \wedge c_2$ also implements $c_1$ and~$c_2$.
Thus, every component in $M_{c_1 \wedge c_2}$ is also
in $M_{c_1}$ and $M_{c_2}$.
Similarly, all components that are environments of
$c_1$ or $c_2$, are also environments of $c_1 \wedge c_2$.
So $M_{c_1 \wedge c_2}$ is the greatest lower bound
of $M_{c_1}$ and $M_{c_2}$ with respect to refinement.

We will now show that Statement~\ref{itm:meta-inst-comp} holds.
Recall Theorem~\ref{thm:contract-composition},
and say that we have components and contracts such that
$m_1 \models c_1$ and $m_2 \models c_2$, or,
by Definition~\ref{def:meta-contract},
that $m_1 \models^M \mathcal{C}_{c_1}$
and $m_2 \models^M \mathcal{C}_{c_2}$.
From Theorem~\ref{thm:contract-composition} we then have that
$m_1 \times m_2 \models c_1 \otimes c_2$, or
$m_1 \times m_2 \models^M \mathcal{C}_{c_1 \otimes c_2}$.
This trivially holds in the other direction as well.
Say that we also have $m$ such that $m$ is an environment
to $c_1 \otimes c_2$, or, in other words, that
$m \models^E \mathcal{C}_{c_1 \otimes c_2}$.
By Theorem~\ref{thm:contract-composition} we then have that
$m_1 \times m$ is an environment to $c_2$, or
$m_1 \times m \models^E \mathcal{C}_{c_2}$.
Again, this trivially holds in the other direction,
and by symmetry we also have that
$m \times m_2 \models^E \mathcal{C}_{c_1}$.
\end{proof}

Let us now return to our example from Section~\ref{sec:preliminaries}.
When applying Contract Based Design, contracts at the
more abstract level will be decomposed into contracts at the
more concrete level. So, for our example, we might have
at the top level a contract $c = (\rho_c^-, \rho_c^+)$ 
with interface $(\varnothing, \{even, odd\})$, where
$\rho_c^- = \varnothing$, and where $\rho_c^+ \in \fenv{\fprov{c}}$
maps $\textit{even}$ to the set of pairs $(s, s')$ such 
that whenever $s(n)$ is non-negative and even, then
$s'(r) = 1$,
and when $s(n)$ is non-negative and odd, then
$s'(r) = 0$,
and maps $\textit{odd}$ in a dual manner. 
This contract could then be decomposed into two contracts
$c_{\mathit{even}}$ and $c_{\mathit{odd}}$, so that
$\rho^+_{c_{\mathit{even}}} (\mathit{even}) \defeq \rho_c^+ (\mathit{even})$ and 
$\rho^-_{c_{\mathit{even}}} (\mathit{odd}) \defeq \rho_c^+ (\mathit{odd})$, and~$c_{\mathit{odd}}$ is analogous. 
Then, we would have
$c_{\mathit{even}} \otimes c_{\mathit{odd}} \preceq c$,
and for any two components $m_{\mathit{even}}$ and
$m_{\mathit{odd}}$ such that
$m_{\mathit{even}} \models c_{\mathit{even}}$ and
$m_{\mathit{odd}} \models c_{\mathit{odd}}$, it would
hold that
$m_{\mathit{even}} \times m_{\mathit{odd}} \models c$.


\section{Connection to Programs with Procedures}
\label{sec:hoare-connection}

In this section we discuss how
our abstract contract theory from
Section~\ref{sec:proc-lang-contract-theory}
relates to programs with procedures as presented in
Section~\ref{subsec:denot-sem},
and how it relates to Hoare logic and
procedure-modular verification
as presented in Section~\ref{subsec:prog-contr}.
%
%



First, we define how to abstract the denotational
notion of procedures into components in the abstract
theory, based on the function $\xi$ from
Section~\ref{subsec:denot-sem}.

\begin{definition}[From procedure sets to components]
\label{def:from-procedure-sets-to-components}
For any set of procedures $\fprov{}$, calling
procedures $P'$, we define
the component $\comptype{m}$, where 
$\freq{m} \defeq P' \setminus \fprov{m}$ and
$\fprov{m} \defeq \fprov{}$, so that
$\forall \rho^-_m \in \fenv{\freq{m}}.\ 
\forall p \in \fprov{m}.\ 
m(\rho^-_m)(p)
\defeq
\denot{S_p}_{\rho^-_m}
$.
\end{definition}

As the next result shows,
procedure set abstraction and component composition commute.
Together with commutativity and associativity of component
composition, this means that the initial grouping of procedures
into components is irrelevant, and that one can start with
abstracting each individual procedure into a component.

\begin{theorem}
\label{thm:abstraction-composition-commutativity}
For 
any two disjoint sets of procedures $\fprov{1}$ and $\fprov{2}$,
abstracted individually into components $m_{1}$ and $m_{2}$, 
respectively, and $\fprov{1} \cup \fprov{2}$ abstracted
into component~$m$, it holds that
$m_{1} \times m_{2} = m$.
\end{theorem}

\begin{proof}
Let the set of procedures~$\fprov{1}$, calling but not
providing $\freq{1}$, and the disjoint set~$\fprov{2}$,
calling but not providing $\freq{2}$, be abstracted
into components $m_{1}$ and~$m_{2}$, respectively,
and let the set $\fprov{} = \fprov{1} \cup \fprov{2}$
be abstracted into component~$m$, all following
Definition~\ref{def:from-procedure-sets-to-components}.
The interfaces of components~$m$ and~$m_1 \times m_2$
are equal, since
$\fprov{m} = \fprov{m_1} \cup \fprov{m_2}
= \fprov{m_1 \times m_2}$,
and if $\freq{2}$ and $\freq{2}$ are the procedures called
by (but not in) $\fprov{1}$ and $\fprov{2}$, respectively,
then the procedures called by (but not in) $\fprov{}$ are
$\freq{m}
= (\freq{m_1} \cup \freq{m_2}) \setminus
    (\fprov{m_1} \cup \fprov{m_2})
= \freq{m_1 \times m_2}$.

To show that $m_1 \times m_2 = m$, we have also to show 
that for all procedures $p \in \fprov{1} \cup \fprov{2}$,
$\forall \rho \in \fenv{\freq{m}}.\ m(\rho)(p)
= (m_1 \times m_2)(\rho)(p)$.
%
%
It is enough to show that
$\forall p \in \fprov{1}.\ 
\forall \rho \in \fenv{\freq{m}}.\ m(\rho)(p)
= (m_1 \times m_2)(\rho)(p)$.
By symmetry it then also holds for all $p \in \fprov{2}$,
and thus for all $p \in \fprov{m}$.

For any $p \in \fprov{1}$ and $\rho^- \in \fenv{\freq{m}}$,
$m(\rho^-)(p) = \denot{S_p}_{\rho^-} = \denot{S_p}_{\rho^-}^{\rho^+_0}
= \rho^+_0(p) = (\mu \rho.\ \xi(\rho))(p)$, where
$\xi : \fenv{\fprov{}} \rightarrow \fenv{\fprov{}}$
is relativised on $\rho^-$, as we recall from
Section~\ref{subsec:denot-sem}.
However, we could view this as taking the
simultaneuos least fixed-point of two functions
$\xi_1 : \fenv{\fprov{1}} \times \fenv{\fprov{2}}
\rightarrow \fenv{\fprov{1}}$ and
$\xi_2 : \fenv{\fprov{1}} \times \fenv{\fprov{2}}
\rightarrow \fenv{\fprov{2}}$,
both also relative to~$\rho^-$, and then taking the union
of the resulting procedure environments as the result.

The well-known Beki\'{c}'s Lemma~\cite{bek-84-lemma} states that,
for a complete lattice $L$, and monotone $f, g$ such that
$f: L^{p+q} \rightarrow L^p$, $g: L^{p+q} \rightarrow L^q$,
it holds that
$\mu x, y.\ (f(x, y), g(x, y)) = (x_0, y_0)$, where
$x_0 = \mu x.\ f(x, \mu y.\ g(x, y))$ and
$y_0 = \mu y.\ g(x_0, y)$,
meaning that for two interdependent monotone functions,
taking the simultaneous least fixed-points gives the same
results as iteratively finding them one function at a time.
Using the special case when $p=q=1$, we have that the
lemma also holds for $f, g : L \times L \rightarrow L$.
Furthermore, in our case, for any complete lattice
$(\fenv{P}, \sqsubseteq)$ and function
$h': \fenv{P} \rightarrow \fenv{P}$,
we can define the equivalent function
$h: \fenv{} \rightarrow \fenv{}$ by
$h(\rho) \defeq h'(\rho \sqcap \rho^\top_{P})$,
essentially ignoring denotations for all procedures
not in $P$.
We then get that Beki\'{c}'s Lemma also holds for monotone
$f: \fenv{P_1} \times \fenv{P_2} \rightarrow \fenv{P_1}$ and
$g: \fenv{P_1} \times \fenv{P_2} \rightarrow \fenv{P_2}$,
for any two sets of procedures $P_1$ and $P_2$.

Again relativised on $\rho^-$, for $p \in \fprov{1}$ and
$\anyprov{\chi}{m_1 \times m_2} :
\fenv{\fprov{}} \rightarrow \fenv{\fprov{}}$ we have that
$(m_1 \times m_2)(\rho^-)(p)
= (\mu \rho.\ \anyprov{\chi}{m_1 \times m_2}(\rho))(p)$,
according to Definition~\ref{def:comp-composition}.
Here, $\anyprov{\chi}{m_1 \times m_2}$ is the simultaneous
fixed-point of two functions, for which we already have
partial fixed-points, of the function $\xi$ from
Section~\ref{subsec:denot-sem}.
For the procedures in $P^+_1$ in particular, and for each
$\rho^-_1 \in \fenv{\freq{1}}$, we have the fixed-points
of the function
$\xi'_1 : \fenv{\fprov{1}} \rightarrow \fenv{\fprov{1}}$.
Recall $\xi_1$ from above.
When $\rho^- \cup \rho^-_2 = \rho^-_1$, we have
$\xi_1(\rho^-_2, x) = \xi'_1(x)$
relative to $\rho^-$ and~$\rho^-_1$, respectively.
Since these functions are all monotone, when taking the
least fixed-point of $\anyprov{\chi}{m_1 \times m_2}$,
using the already computed fixed-points, we will then
by Beki\'{c}'s Lemma get the same result as the least
fixed-point of $\xi$ above, and thus $m = m_1 \times m_2$.
\end{proof}

\paragraph{Component abstraction example.}

Let us illustrate the theorem on our even-odd  example
(however, the example does not really illustrate Beki\'{c}'s 
Lemma, since the two procedures do not call themselves).

By Definition~\ref{def:from-procedure-sets-to-components}, 
the procedure set $\set{\mathit{even}}$ 
is abstracted into component 
$m_{\mathit{even}} : \fenv{\{odd\}} \rightarrow \fenv{\{even\}}$
with interface $(\{odd\}, \{even\})$, so that
$\forall \rho^- \in \fenv{\{odd\}}.\  
  m(\rho^-)(\mathit{even}) = \denot{S_{\mathit{even}}}_{\rho^-}$.
By definition, $\denot{S_{\mathit{even}}}_{\rho^-}$
is equal to 
$\denot{S_\mathit{even}}_{\rho^-}^{\rho_0^+}$, where
$\rho_0^+$ is the least fixed point of 
$\xi : \fenv{\set{\mathit{even}}} \rightarrow \fenv{\set{\mathit{even}}}$
defined by
$\xi (\rho^+) (\mathit{even}) \defeq 
 \denot{S_\mathit{even}}_{\rho^-}^{\rho^+}$
for any $\rho^+ \in \fenv{\{even\}}$. 
Notice, however, that procedure $\mathit{even}$ does 
not have any calls to itself, so
$\denot{S_\mathit{even}}_{\rho^-}^{\rho_0^+}$
does not really depend on~$\rho^+$. 
Then, for any $\rho^- \in \fenv{\{odd\}}$,
$(s, s') \in  m(\rho^-)(\mathit{even})$
if either $s(n)=0$ and $s' = s[r \mapsto 1]$, or else if
$s(n)>0$ and
$(s[n \mapsto s(n)-1], s') \in \rho^- (\mathit{odd})$.

Similarly, the procedure set $\set{\mathit{odd}}$ 
is abstracted into component 
$m_{\mathit{odd}} : \fenv{\{even\}} \rightarrow \fenv{\{odd\}}$
with interface $(\{even\}, \{odd\})$, so that
$\forall \rho^- \in \fenv{\{even\}}.\ $ 
$m(\rho^-)(\mathit{odd}) = \denot{S_{\mathit{odd}}}_{\rho^-}$.
Then, for any $\rho^- \in \fenv{\{even\}}$,
$(s, s') \in  m(\rho^-)(\mathit{odd})$
if either $s(n)=0$ and $s' = s[r \mapsto 0]$, or else if
$s(n)>0$ and
$(s[n \mapsto s(n)-1], s') \in \rho^- (\mathit{even})$.

Now, applying 
Definition~\ref{def:from-procedure-sets-to-components}
to the whole (closed) program yields a component
$m : \fenv{\varnothing} \rightarrow 
     \fenv{\set{\mathit{even}, \mathit{odd}}}$
with interface
$(\varnothing, \set{\mathit{even}, \mathit{odd}})$,
so that
$\forall \rho^- \in \fenv{\varnothing}.\ 
 \forall p \in \set{\mathit{even}, \mathit{odd}}.\ 
   m(\rho^-)(p) = \denot{S_p}_{\rho^-}$.
Recall the denotations $\denot{S_\mathit{even}}_{\rho^-}$
and $\denot{S_\mathit{odd}}_{\rho^-}$
from the end of Section~\ref{subsec:denot-sem}. 

Components~$m_{\mathit{even}}$ and~$m_{\mathit{odd}}$ are
composable, and by Definition~\ref{def:comp-composition}, 
their composition has (the same) interface
$(\varnothing, \set{\mathit{even}, \mathit{odd}})$,
and is (also) a mapping
$m_\mathit{even} \times m_\mathit{odd} : 
   \fenv{\varnothing} \rightarrow 
   \fenv{\set{\mathit{even}, \mathit{odd}}}$.

Finally, note that function
$\anyprov{\chi}{m_\mathit{even} \times m_\mathit{odd}} : 
   \fenv{\set{\mathit{even}, \mathit{odd}}} \rightarrow 
   \fenv{\set{\mathit{even}, \mathit{odd}}}$
is exactly the function~$\xi$ in the context of the
interface $(\varnothing, \set{\mathit{even}, \mathit{odd}})$.
This can be seen by first noting that since
$\fenv{\varnothing} = \varnothing$, we have that
$\anyprov{\chi}{\excomp}$ only depends on its arguments.
Furthermore, for all
$\rho^+ \in \fenv{\set{\mathit{even}, \mathit{odd}}}$,
if $\rho^+_{\mathit{odd}} \defeq \rho^+_{\big|\set{odd}}$
and $\rho^+_{\mathit{even}} \defeq \rho^+_{\big|\set{even}}$
we have that, since
$\mathit{odd} \in \freq{\mathit{even}} \cap \fprov{\mathit{odd}}$,
then
$\anyprov{\chi}{\excomp}(\rho^+)(\mathit{even})
= m_\mathit{even}(\rho^+_{\mathit{odd}})(\mathit{even})
= \denot{S_\mathit{even}}_{\rho^+_{\mathit{odd}}}
= \denot{S_\mathit{even}}^{\rho^+}
= \xi(\rho^+)(\mathit{even})$.
Similarly $\anyprov{\chi}{\excomp}(\rho^+)(\mathit{odd})
= \xi(\rho^+)(\mathit{odd})$.
We therefore have
$m_\mathit{even} \times m_\mathit{odd} = m$. \\

We now define how to abstract Hoare
logic contracts into denotational contracts,
in terms of the contract environment~$\rho_c$ defined 
in Section~\ref{subsec:prog-contr}.

\begin{definition}[From Hoare logic contracts to denotational contracts]
\label{def:from-hoare-to-denotational-contracts}
For a procedure~$p$ with Hoare logic contract~$C_p$,
calling other procedures~$P^-$,
we define the denotational contract
$c_p = (\rho^-_{c_p}, \rho^+_{c_p})$ with interface
$\fprov{c_p} \defeq \{ p \}$ and $\freq{c_p} \defeq P^-$, so that
$\rho^+_{c_p}(p) \defeq \rho_c(p)$, and 
$\forall p' \in P^-.\ \rho^-_{c_p}(p') = \rho_c(p')$.
\end{definition}

In this way, conceptually, denotational contracts become
assume/guarantee-style specifications over Hoare logic 
procedure contracts: assuming that all (external) procedures
called by a procedure~$p$ transform the state according to 
their Hoare logic contracts, procedure~$p$ obliges itself to
do so as well. 

We now show that if a procedure implements a
Hoare logic contract, then the abstracted component
will implement the abstracted contract,
and vice versa.
Together with Theorem~\ref{thm:abstraction-composition-commutativity},
this result allows the \emph{procedure-modular verification} 
of abstract components.

\begin{theorem}
\label{thm:abstraction-contract-commutativity}
For any procedure~$p$ with procedure contract~$C_p$,
abstracted into component~$m_p$ with contract~$c_p$,
we have
$S_p \models_{\mathit{par}}^{\mathit{cr}} C_p$
iff $m_p \models c_p$.
\end{theorem}

\begin{proof}
First, the interfaces will agree:
$\fprov{m_p} = \{ p \} = \fprov{c_p}$,
and $\freq{m_p} = \freq{c_p}$ since they
are both the set of called procedures.

Next, $\denot{S_p}^{cr} = \denot{S_p}_{\rho_c}$ by definition
(see Section~\ref{subsec:prog-contr}),
i.e., the contract-relative denotation of~$S_p$
is relativised on the contract
environment~$\rho_c$, and so is~$\rho^-_{c_p}$
according to Definition~\ref{def:from-hoare-to-denotational-contracts}.
If we have that $S_p \models_{\mathit{par}}^{\mathit{cr}} C_p$,
then $\denot{S_p}_{\rho_c} \subseteq \denot{C_p}$,
and since $\rho^+_{c_p}(p) = \denot{C_p}$ then
$\denot{S_p}_{\rho_c} \subseteq \rho^+_{c_p}(p)$.
Because $\rho_c$ and $\rho^-_{c_p}$ agree on all
denotations for procedures in $\freq{c_p}$,
and because only the denotations of the procedures
in $\freq{c_p}$ affect the denotation of $p$, then
we also have
$\denot{S_p}_{\rho^-_{c_p}}
\subseteq \rho^+_{c_p}(p)$.

Since
$m(\rho^-_{c_p})(p) =\denot{S_p}_{\rho^-_{c_p}}$
and
$\fprov{m_p} = \{ p\}$, then
$m(\rho^-_{c_p}) \sqsubseteq \rho^+_{c_p}$.
Finally, since $\freq{m_p} =\freq{c_p}$, we have
$m(\rho_{c_p}^- \sqcup \rho^\top_{\freq{m_p} \setminus \freq{c_p}})
= m(\rho^-_{c_p})$ and therefore $m \models c$.
The same argument holds in the other direction.
\end{proof}


Returning to the example from Sections~\ref{sec:preliminaries}
and~\ref{sec:meta-connection}, we can abstract the
procedure set $\{even\}$ into component $m_{\mathit{even}}$,
with interface $(\{odd\}, \{even\})$, which would be a function
$\fenv{\{odd\}} \rightarrow \fenv{\{even\}}$, and
$\forall \rho^- \in \fenv{\{odd\}}.\  m(\rho^-)(\mathit{even}) =
\denot{S_{\mathit{even}}}_{\rho^-}$.
The denotational contracts $c_{\mathit{even}}$
and $c_{\mathit{odd}}$ resulting from
the decomposition shown in Section~\ref{sec:meta-connection}, 
would be exactly the abstraction of the Hoare Logic contracts
$C_{\mathit{even}}$ and $C_{\mathit{odd}}$
shown in Section~\ref{subsec:prog-contr}.
They would both be part of the contract environment
used in procedure-modular verification, for example
when verifying that
$S_\mathit{even} \models_{\mathit{par}}^{\mathit{cr}} C_\mathit{even}$,
which would entail $m_{\mathit{even}} \models c_{\mathit{even}}$.
Thus, by applying standard procedure-modular verification at
the source code level, we prove the top-level contract~$c$
proposed in Section~\ref{sec:meta-connection}. 


\section{Conclusion}
\label{sec:conclusion}

We presented an abstract contract theory for procedural
languages, based on denotational semantics. The theory
is shown to be an instance of the meta-theory 
of~\cite{ben-et-al-18}, and at the same time an abstraction
of the standard denotational semantics of procedural
languages. 
We believe that our contract theory can be used to support
the development of cyber-physical and embedded systems by
the design methodology supported by the meta-theory, allowing
the individual procedures of the embedded software
to be treated as any other system component.
The work also strengthens the claims of the meta-theory
of distilling the notion of contracts to its essence,
by showing that it is applicable also in the context
of procedural programs and deductive verification.
Finally, this work serves as a preparation for combining
our contract theory for procedural programs with other
instantiations of the meta-theory.
In future work we plan to investigate the utility of our
contract theory on real embedded systems taken from the
automotive industry, where not all components are
procedural programs, or even software
(cf.\ our previous work, e.g., \cite{gur-lid-nyb-wes-17-fmics}).
We also plan to extend our toy imperative language
with additional features,
such as procedure parameters and return values.
Furthermore, we plan to extend the contract theory to
capture program traces by developing a finite-trace
semantics, to enable its use in
the specification and verification of temporal properties.
Lastly, we plan to combine our contract theory with a contract
theory for hybrid systems~\cite{nyb-wes-gur-20-isola}.


\bibliographystyle{splncs03}
\bibliography{refs}

\end{document}